\documentclass[12pt]{article}
 \usepackage{mathtools}
\usepackage{amssymb}
\usepackage{amsbsy}
\usepackage{epsfig}
\usepackage{verbatim}
\usepackage[usenames,dvipsnames]{color}

\makeatletter \@addtoreset{figure}{section}
\def\thefigure{\thesection.\@arabic\c@figure}
\def\fps@figure{h, t}
\@addtoreset{table}{bsection}
\def\thetable{\thesection.\@arabic\c@table}
\def\fps@table{h, t}
\@addtoreset{equation}{section}

\newtheorem{corollary}{Corollary}[section]
\newtheorem{definition}{Definition}[section]
\newtheorem{theorem}{Theorem}[section]
\newtheorem{proposition}{Proposition}[section]

\newtheorem{lemma}{Lemma}[section]
\newtheorem{remark}{Remark}[section]
\newtheorem{remarks}[remark]{Remarks}

\def\bd{\begin{definition}}
\def\ed{\end{definition}}
\def\bt{\begin{theorem}}
\def\et{\end{theorem}}
\def\bp{\begin{proposition}\rm}
\def\ep{\end{proposition}}
\def\bc{\begin{corollary}}
\def\ec{\end{corollary}}
\def\bl{\begin{lemma}\em}
\def\el{\end{lemma}}
\def\be{\begin{equation}}
\def\ee{\end{equation}}
\def\br{\begin{remark}\rm\small}
\def\er{\end{remark}}
\def\brs{\begin{remarks}.\\ \rm\
\begin{enumerate}}
\def\ers{\end{enumerate}\end{remarks}}
\def\bea{\begin{eqnarray}}
\def\eea{\end{eqnarray}}

\def\ra{{\rightarrow}}

\def\det{\mathrm {det}}

\def\span{\mathrm {span}}

\def\diag{\mathrm {diag}}
\def\span{\mathrm {span}}

\def\Gr{\mathrm {Gr}}

\def\&{&{\hskip -20pt}}

\def\HH{\mathcal{H}}

\def\OO{\mathcal{O}}

\def\Cb{\mathbf{C}}

\def\Fb{\mathbf{F}}

\def\Ib{\mathbf{I}}

\def\Nb{\mathbf{N}}

\def\Zb{\mathbf{Z}}

 \def\grl{\mathfrak{l}}

\def\grP{\mathfrak{P}}

\def\nchi{\hbox{\raise 2.5pt\hbox{$\chi$}}}
\date{}
\begin{document}
\baselineskip 16pt 
\begin{flushright} 
CRM-3326 (2013)
 \end{flushright}
\medskip
\begin{center}
\begin{Large}\fontfamily{cmss}
\fontsize{17pt}{27pt}
\selectfont
\textbf{Symmetric polynomials, generalized Jacobi-Trudi identities and $\tau$-functions}\footnote{Work of J.H. supported by the Natural Sciences and Engineering Research Council of Canada (NSERC) and the Fonds Qu\'ebecois de la recherche sur la nature et les technologies (FQRNT). }
\end{Large}\\
\bigskip
\begin{large}  {J. Harnad}$^{1,2}$
 and {Eunghyun Lee}$^{1,2}$
 \end{large}
\\
\bigskip
\begin{small}
$^{1}${\em Centre de recherches math\'ematiques,
Universit\'e de Montr\'eal\\ C.~P.~6128, succ. centre ville, Montr\'eal,
Qu\'ebec, Canada H3C 3J7 } \\
\smallskip
$^{2}${\em Department of Mathematics and
Statistics, Concordia University\\ 1455 de Maisonneuve Blvd. W. 
Montreal, Quebec,  Canada H3G 1M8 } 
\end{small}
\end{center}
\bigskip

\begin{center}{\bf Abstract}
\end{center}
\smallskip

\begin{small}

 An element $[\Phi] \in Gr_n (\HH_+, \Fb)$ of the Grassmannian of $n$-dimensional subspaces of the 
Hardy space $ \HH_+=H^2$,  extended over the field $\Fb=\Cb(x_1,\dots, x_n)$, may be
associated to any polynomial basis  $\phi = \{\phi_0, \phi_1, \cdots\} $ for $\Cb(x)$. The Pl\"ucker coordinates $S^\phi_{\lambda,n}(x_1,\dots, x_n)$ of $[\Phi]$,  labelled by partitions $\lambda$,  provide an analog of  Jacobi's  bi-alternant  formula,
defining a generalization of  Schur polynomials.
Applying the recursion relations satisfied by the polynomial system $\phi$  to the analog 
$\{h^{(0)}_i \}$ of the complete symmetric functions  generates 
a doubly infinite matrix $ h^{(j)}_i$  of symmetric polynomials that determine an element $[H]\in \Gr_n(\HH_+, \Fb)$.
This is shown to coincide with $[\Phi]$, implying a set of  generalized Jacobi identities, extending a result obtained by Sergeev and Veselov \cite{SV} for the case of orthogonal polynomials.
  The symmetric polynomials $S^\phi_{\lambda,n}(x_1,\dots, x_n)$ are shown to be KP (Kadomtsev-Petviashvili) 
   $\tau$-functions in terms  of  the monomial sums $[x]$ of the $x_a$'s,   viewed as KP flow variables. A fermionic
 operator representation is derived for these, as well as for the infinite sums  $ \sum_{\lambda}S_{\lambda,n}^\phi([x]) S^\theta_{\lambda,n} ({\bf t})$
associated to  any pair of polynomial bases $(\phi, \theta)$, which are shown to be  2D Toda lattice  $\tau$-functions.
A number of applications are given, including classical group character expansions,
matrix model partition functions and generators for random processes.
    \bigskip
\end{small}
\bigskip \bigskip

\section{Introduction }

The Jacobi-Trudi identities \cite{Mac}  express   Schur polynomials 
$S_\lambda(x_1, x_2, \dots, x_n)$,  labelled  by partitions $\lambda$  of 
length $\ell(\lambda)\le n$,  as  determinants 
\be
S_\lambda = \det (h_{\lambda_i -i +j}) = \det(e_{\lambda'_i -i +j})
\ee
in terms of the complete and elementary symmetric functions
\be
h_j := S_{(j)},  \quad e_j := S_{(1^j)}.
\ee
Here $\lambda =(\lambda_1 \ge \lambda_2, \dots \ge \lambda_{\ell(\lambda)}> 0 ) $ denotes a partition 
and $\lambda'$ the conjugate partition, whose Young diagram  is the transpose
of the one for $\lambda$.

In recent years, a number of examples have been found in which  a generalized form of such
identities are satisfied by  other classes of functions  \cite{Mac2, Ch, BR, Ku, AKLTZ, SV}, also labelled by partitions,
together with an additional integer parameter. These are referred to as  
generalized, or ``quantum''  Jacobi-Trudi identities. In particular, they are known to be 
satisfied by the eigenvalues of  commuting transfer matrices  in the $R$-matrix approach to quantum 
integrable systems, due to the underlying Yangian algebra structure, and are known in this case 
as the Bazhanov-Cherednik-Reshetikhin  formula   \cite{Ch, BR, AKLTZ}. 
 
  Another recently studied case  \cite{SV} involves a generalization $S^\phi_{\lambda,n}$
 of  Schur functions obtained by replacing the monomials appearing in
 the bi-alternant  formula  for   $S_\lambda$ by a sequence of orthogonal polynomials 
 $\{\phi_i\}_{i\ =0, 1,\dots}$.
  In this case, the additional integer labels  a sequence of functions
 $\{h^{(j)}_i\}_{i,j \in \Nb}$ defined recursively in terms of the analogs  $h_i^{(0)} := S^\phi_{(i),n}$ of
 the complete symmetric functions.
More generally, it has been shown \cite{AKLTZ}, that such generalized Jacobi-Trudi identities are satisfied 
by the coefficients of the expansion of any  sequence of KP $\tau$-functions determining a solution of the MKP integrable hierarchy in a basis
 of Schur functions.
 
 In the following, we extend the results of ref.~\cite{SV} to the case of arbitrary
 polynomials bases  $\phi:= \{\phi_i\}_{i\ =0, 1,\dots}$ of $\Cb(x)$, using a geometrical approach 
 that associates to any such polynomial system a corresponding 
 element  $[\Phi]$ of an infinite Grassmann manifold, whose Pl\"ucker coordinates coincide
 with the functions $S^\phi_{\lambda, n}$. The meaning of the Jacobi-Trudi identities becomes
clear in this setting;  they follow from the fact that the element $[H]$ of the Grassmannian 
 determined from  applying the same set of recursions  relations as those satisfied by the polynomials
 $\{\phi_i\}_{i\ =0, 1,\dots}$ to the analogs   $\{h^{(j)}_i\}_{i,j \in \Nb}$  of the complete symmetric functions is, 
 within a change of basis, identical to the element $[\Phi]$ determined
 by evaluation of the polynomials themselves at the various parameter values
 $\{x_1, \dots , x_n\}$, and hence their Pl\"ucker coordinates coincide. This result is 
 the content of Proposition \ref{H_Phi}, Section 2.2, with a dual version given in Section 2.3.  
 Two methods of proof are given:  induction in the successive basis elements defining $[H]$, 
 and  the ``dressing'' method, based on transforming the standard case, involving monomials systems, 
 to the general one.
 
In Section 3, it is shown that the resulting functions $S^\phi_{\lambda,n}([x])$ may be viewed as KP
 $\tau$-functions in terms of flow variables identified  as monomial sums $[x]$ over
the parameters $\{x_1, \dots, x_n \}$.  Moreover,  since the symmetric polynomials $S^\phi_{\lambda,n}([x]_)$
 also satisfy the Pl\"ucker relations,  they may be  used as coefficients in a Schur function series
defining a parametric  family of KP $\tau$-functions
 \be
 \tau_\phi(n, [x] ,  {\bf t}) =\sum_{\lambda \atop \ell(\lambda)\le n} S^\phi_{\lambda,n}(x_1, \dots , x_n)  S_\lambda({\bf t}),
 \ee
 where the variables ${\bf t}=(t_1, t_2, \dots)$ are viewed as additional KP flow parameters.
 A fermionic representation of these $\tau$-functions is  given, and used
 to show that they form a lattice of $\tau$-functions in the ${\bf t}$  variables which, 
 when combined with the dependence on the monomials sum variables  $[x]$, may
 also be viewed as a 2D-Toda $\tau$-functions.
More generally, summing the products of generalized Schur functions $S^\phi_{\lambda,n}$ and $S^\theta_{\lambda,n}$
corresponding to a pair $(\phi, \theta)$ of such polynomials systems 
 \be
 \tau_{\phi, \theta}(n, {\bf s}, {\bf t}) = \sum_{\lambda} S^\phi_{\lambda,n} ({\bf s})S^\theta_{\lambda,n} ({\bf t})
 \ee
 provides a broader class of 2D-Toda $\tau$-functions.
 
    In Section 4, these results are iapplied to various examples, including:
determinantal representations of irreducible characters of the classical Lie groups
and their  Schur function expansions; generalized matrix model 
partition functions and generating functions for certain exclusion processes.

\section{Polynomial systems and Grassmannians }

\subsection{Generalized Schur polynomials $S^\phi_{\lambda}$}

 Let $\Fb:= \Cb(x_1,  \dots, x_n)$ denote the extension of the field of complex numbers by the  indeterminates $\{ x_1, x_2, \dots , x_n\}$.  Let $\HH_+$ denote the space of those square integrable  complex  functions  on the unit circle $|z|=1$  that admit a holomorphic extension to the interior  disc
 (i.e., the Hardy space $H^2$). We  use the   monomial basis $\{{\bf b}_i  := z^{i-1}\}_{i\in \Nb^+}$ to represent elements as semi-infinite column vectors, labelled increasingly from the bottom element  upward; i.e. ,
 ${\bf b}_1 \sim  (\dots , 0, 0,  1)^t$,  ${\bf b}_2 \sim  (\dots ,  0,  1, 0)^t$, etc. 
 
  Denote by $\Gr_n(\ \HH_+, \Fb)$ the Grassmannian  of $n$-dimensional subspaces of  
  \be
\HH_+(\Fb):=   \Fb\otimes_{\Cb} \HH_+.
  \ee
  Relative to the basis $\{{\bf b}_i\}$, an element  $[W] \in \Gr_n( \HH_+, \Fb)$ may be represented by a semi-infinite $(\infty \times n)$  rank $n$ matrix $W$ whose columns $(W^1, \dots , W^n)$ span $[W]$. 
 Two such rank-$n$ matrices related by $W = \tilde{W}g$, $g\in Gl(n, \Fb)$, span the same subspace,  and hence belong to the same equivalence class  $[W]= [Wg]$. The $n$-component row vectors of  $W$, corresponding to the components along the basis elements $b_i$ will be denoted  
\be
 W_i:= (W_{i1}, \quad \dots W_{in}), \quad i\in \Nb. 
 \ee

For any integer partition D
\be
\lambda= (\lambda_1 \ge \lambda_2 \ge \dots \ \lambda_{\ell(\lambda)} \ge 0),  \quad \lambda_i \in \Nb^+
\ee
of weight
\be
|\lambda| = \sum_{i=1}^{\ell(\lambda)} \lambda_i
\ee
 and length $\ell(\lambda)$, we define an infinite, strictly decreasing sequence of integers  $l_1 > l_2 > \dots $
 (sometimes called ``particle coordinates'') by
\be
l_i := \lambda_i - i +n ,  \quad i \in \Nb^+, 
\label{part_coords}
\ee
with the convention that $\lambda_i:=0$ for $i> \ell(\lambda)$.
After $\ell(\lambda)$ terms, these   become the successive  decreasing integers
 $(n -\ell(\lambda), n-\ell(\lambda)-1, \dots)$.
For any partition of length
\be
\ell(\lambda) \le n
\ee
let $W_\lambda$ denote the $n \times n$ minor of $W$ whose $i$th row (counting from the
top down) is $W_{l_i+1}$.
Then the $\lambda$th Pl\"ucker coordinate of the  image of the element $W$ under the Pl\"ucker map
\be
\grP\grl: \Gr_n(\HH_+, \Fb) \ra \bigwedge^n \HH_+(\Fb)
\ee
is given, within projective equivalence, by the determinant
\be
\pi_\lambda(W) := \det(W_\lambda).
\ee
These satisfy the Pl\"ucker relations which, on the ``big cell'' (i.e. where $\pi_0(W)\ne 0$),
are equivalent to the generalized Giambelli identity  (see e.g, \cite{HE}, Corollary 2.1):
\be
\pi_0(W)^{r-1}\pi_\lambda(W) = \det (\pi_{(a_i | b_j)})  |_{1\le i,j \le r}.
\label{gen_Giambelli}
\ee
Here $(a_1, \dots , a_k | b_1, b_2, \dots b_r)$ is  Frobenius' notation for the partition $\lambda$, with $(a_i, b_i)$ 
the arm and leg lengths of the Young diagram for $\lambda$, measured from the $i$th diagonal, $(a_i|b_j)$ denotes the hook partition 
$\lambda = (1+a_i, 1^{b_j})$ and $r$ is the Frobenius rank of $\lambda$ (i.e., the length of the central diagonal).

   Given a basis for $\Cb[x]$ consisting of monic polynomials $\{\phi_i(x), \ \deg \phi_i=i\}_{i \in \Nb}$ , we associate an 
   $\infty \times n$ matrix $\Phi$  with components 
   \be
   \Phi_{ij}:= \phi_{i-1}(x_j)
   \ee
   that determines an element $[\Phi] \in \Gr_n(\HH_+^n, \Fb)$.
  The  $i$th row  vector is thus
\be
\Phi^t_i: = (\phi_{i-1}(x_1), \dots , \phi_{i-1}(x_n)).
\ee
Let $\Lambda$ denote the semi-infinite upper triangular  shift matrix   (with $1$'s above the principal diagonal and
zeros elsewhere), representing multiplication by $z$: 
\be
z: {\bf b}_i \ra {\bf b}_{i+1}.
\ee
and  $\Lambda^t$ its (lower triangular) transpose, and define
\be
X := \diag(x_1, \dots x_n).
\ee
It follows that there is a unique semi-infinite upper triangular recursion matrix $J^+$ such that
\be
\Phi X = J\Phi 
\label{recurs_Phi}
\ee
where
\be 
J := \Lambda^t + J^+.
\ee
(In the special case where the $\phi_i$'s form an orthonormal system with respect to some inner product, 
the matrix $J$ is tridiagonal, but this will not be assumed here.)

Define the infinite triangular  matrix $A^\phi$, with $1$'s along the diagonal, whose rows are the coefficients of 
the polynomials $\{\phi_i\}$, 
\be
A^\phi_{ij}:= \phi_{i-1, j-1} \quad {\rm if } \quad i\ge j, \quad A^\phi_{ij} =0 \quad {\rm if } \quad i < j, \quad i,j \in \Nb^+,
\ee
where
\be
\phi_i (x) = \sum_{j=0}^i\phi_{i, j} x^j.
\ee
(Note that  in our notational conventions this is {\it upper} triangular, because we count upward from the
bottom, starting with 1.)
For the case of monomials, denote the matrix $\Phi$  as $\overset{\scriptscriptstyle 0}{\Phi}$.
We then have
 \be
 \Phi = A^\phi \overset{\scriptscriptstyle 0}{\Phi}. 
 \ee
 Since $\overset{\scriptscriptstyle 0}{\Phi}$ satisfies the recursion relations
 \be
 \overset{\scriptscriptstyle 0}{\Phi}X = \Lambda^t  \overset{\scriptscriptstyle 0}{\Phi}
 \ee
 we have the intertwining relation
 \be
 A^\phi \Lambda^t = J A^\phi.
 \label{A_Phi_J_A}
 \ee
Since the infinite matrix $A^\phi$ is the sum of the identity matrix $\Ib_\infty$ 
and a strictly upper triangular one,
its inverse exists (and has elements that are polynomials in those of $A^\phi$). 

We may therefore solve (\ref{A_Phi_J_A}) for the recursion matrix $J$ 
\be
J = A^\phi \Lambda^t \left( A^\phi\right)^{-1}.
\ee
In what follows, it will also be useful to define
\be
\tilde{J}:= A^\phi \Lambda \left(A^\phi\right)^{-1},
\ee
which is a right inverse of $J$
\be
J \tilde{J} = \Ib_\infty,
\ee
but not quite a left inverse
\be
\tilde{J}J = \Ib_\infty - {\bf a} {\bf b}_1^t.
\ee
Here ${\bf b}_1^t$ is the seminfinite unit row vector $( \cdots, 0, 0, 1)$
and ${\bf a}$ is the right-most column vector of $A^\phi$
\be
{\bf a} = A^\phi {\bf b}_1.
\ee

For $k  \in \Nb$, define the $n$-dimensional projection operator
\bea
&\Pi_k: \Fb\otimes_\Cb \HH_+ &{\hskip -10pt}\ra \ \span \{{\bf b}_1,  \dots {\bf b}_n\}  \cr
& \Pi_k : {\bf b}_{j+k} &{\hskip -10pt} \mapsto {\bf b}_{j}  \quad {\rm if} \quad  1 \le j \le n\cr
& \Pi_k : {\bf b}_{j+k} &{\hskip -10pt} \mapsto {\bf 0}  \quad {\rm \quad \ otherwise}.
\eea
In the $\{{\bf b}_i\}$ basis, this is represented by the  semi-infinite $n \times \infty  $ matrix  $\Gamma_k$ 
whose column vectors  all vanish, except in the successive (descending, according to our labelling)  
positions: $(n+k, \dots k+1)$, where they form the $n \times n$ identity matrix $\Ib_n$.

 The projection of the  $\infty \times n$ matrix $W$ onto its $k$th $n\times n$ block is
 denoted
 \be
 W(k) := \Gamma_k W = W_{(k^n)}
 \ee
 (where the RHS expresses the fact that this is just the $n \times n$ minor corresponding 
 to the $n\times k$ rectangular partition $(k^n)$).
 Similarly, the restriction and projection of the $\infty \times \infty$ matrix J to the $k$th $n$-dimensional subspace
 is given by the $n\times n$ matrices
 \be
 J(k):= \Gamma_k J \Gamma_k^t = \Lambda_n^t + J^+(k)
 \ee
 where
 \be
  J^+(k) := \Gamma_k J^+ \Gamma_k^t 
 \ee
 and $\Lambda_n^t$ is the $n\times n$ lower triangular shift matrix.
 
 Applying this to $\Phi$, we obtain
 \be
 \Phi(k) = \begin{pmatrix} \phi_{n+k -1}(x_1) &  \phi_{n+k -1}(x_2) & \cdots & \phi_{n +k -1}(x_n) \cr
                                           \vdots          &  \vdots  & \vdots  & \vdots   \cr     
                                            \phi_{k+1 }(x_1) &  \phi_{k+1 }(x_2) & \cdots & \phi_{k+1}(x_n) \cr      
                                           \phi_{k}(x_1) &  \phi_{k}(x_2) & \cdots & \phi_{k}(x_n)
                                           \end{pmatrix}. 
  \ee
In particular
\be
 \Phi(0) = \begin{pmatrix} \phi_{n -1}(x_1) &  \phi_{n -1}(x_2) & \cdots & \phi_{n  -1}(x_n) \cr
                                           \vdots          &  \vdots  & \vdots  & \vdots   \cr 
                                            \phi_{1}(x_1) &  \phi_{1}(x_2) & \cdots & \phi_{1}(x_n) \cr          
                                           1 &  1 & \cdots & 1 
                                            \end{pmatrix}.
  \ee
  Note that,  since all $\phi_k$'s are monic, the determinant of $ \Phi(0)$ is just the Vandermonde determinant
  \be
  \det (\Phi(0)) = \Delta(x_1, \dots x_n) = \prod_{i< j} (x_i - x_j).
  \ee

  Projecting onto the $k$th $n$-dimensional subspace, the recursion relations (\ref{recurs_Phi})
  give the following $n$-fold sequence of  relations, valid for all  $k \in \Nb$
  \be
  \Phi(k) X = J^+(k) \Phi(k) + \Phi(k+ 1),  \quad  k=0, 1, 2, \dots.
  \label{rec_rels_Phi_k}
  \ee

Now define, as in \cite{SV}, the following generalizations of the usual  Schur functions $S_\lambda(x_1, \dots, x_n)$,
\be
S_\lambda^{\phi} (x_1, \dots , x_n) := {  \det(\Phi_\lambda) \over \det (\Phi(0))} = {\pi_\lambda(\Phi) \over \pi_{0} (\Phi)}.
\label{Sphi_def}
\ee
\br
The classical case, for which (\ref{Sphi_def}) becomes the usual bi-alternant formula  of Jacobi, corresponds to choosing $\phi_i(x) := x^i$, for which $J^+=0$.
and $J=\Lambda^t$. In \cite{SV}, the case of orthogonal polynomials, for which the recursion matrix is $J$ is tridiagonal, was considered. No such restriction is needed in the following;  the results hold for arbitrary polynomial systems.
\er

From the generalized Giambelli identity (\ref{gen_Giambelli}) and the fact that the $S^\phi_\lambda$ are Pl\"ucker cordinates
of the element $[\Phi]$, with $S^\phi_0(x_1, x_2 \dots, x_n) =1$, we have the Giambelli identity for generalized Schur functions
\be
S^\phi_\lambda = \det \left( S^\phi_{(a_i | b_j)}\right) |_{1 \le i,j \le \ell(\lambda)},
\ee
where $(a_1  \dots a_k | b_1 \dots b_k)$  is $\lambda$ in Frobenius notation.

\subsection{The generalized Jacobi-Trudi formula: first form}

The analogs of the {\it complete symmetric functions} are  denoted 
\bea
\label{h0i}
h^{(0)}_i &\& := S^\phi_{(i)} \quad {\rm for}  \quad  i\ge 0, \\
\label{h0_-i}
  h^{(0)}_{-i} &\& :=0 \quad 1\le i \le n-1. 
\eea
We may view these as the components of the column $H^{(1)}$ of an $\infty \times \infty $ matrix ${\bf H}$, 
whose elements are denoted
\be
H_{ij} := h^{(j-1)}_{i-n},  \quad i, j \in  \Zb,  \quad -\infty < j \le n, \quad 1 \le i \le \infty.
\ee
The  labelling conventions are such that the $j$th column is $H^{(j)}$, with 
 $j$  increasing  from left to right  consecutively from $-\infty$ to $n$.
The components of  the column vector $H^{(j)}$ are thus
\be
H^{(j)}_i := H_{ij}, \quad 1\le i \le \infty
\ee
with the first column given by
\be
H^{(1)}_i = h^{(0)}_{i-n}.
\ee
The rows are labelled consecutvely, with $i$ increasing upward from  $1$ to $\infty$, starting at the bottom.
The successive columns $ H^{(2)} \dots H^{(n)} $ are defined from the  $H^{(1)}$ using the same recursion relations
as those satisfied by  the polynomial sequence $\{\phi_i\}_{i\in \Nb}$
\be
H^{(j)} := J H^{(j-1)} = J^{j-1} H^{(1)}, \quad 1 \le j \le n.
\label{rrHj}
\ee
Equivalently,
\be
h^{(j+1)}_{i-n} = h^{(j)}_{i+1-n}  + \sum_{k=1}^i J_{ik}h^{(j)}_{k-n} 
\label{hij}
\ee
The columns $H^{(j)}$ with $j\le 0$ are also defined so 
the recursion relations (\ref{rrHj}) hold
\be
H^{(j-1)} := \tilde{J}H^{(j)}
\ee
Multiplying on the left by $J$  shows that the recursion relation (\ref{rrHj})
also holds for  $j\le 0$.
It  follows  from (\ref{h0_-i}) that
\be
h^{(j)}_{-j} = 1, \quad  h^{(j)}_{-k} =0 \quad {\rm if} \quad  k > j .
\label{negative_h0}
\ee

Let $H$ denote the $\infty \times n$ matrix consisting of the first $n$ columns of ${\bf H}$:
\be
H := \left(H^{(1)} \ H^{(2)} \dots H^{(n)}\right).
\ee
As before, denote by $H(k)$ the $n\times n$ minor obtained by projecting onto the
$n$-dimensional subspace $\span({\bf b}_{n+k-1}, \dots {\bf b}_k)$
\be
H(k) := \Gamma_k H = H_{(k^n)},  \quad k=0, 1, \dots
\ee
and let $H^{(j)}(k)$ denote its $j$th column vector.
The recursion relations (\ref{rrHj}) can equivalently be written
\be
H^{(j+1)}(k) = J^+(k)H^{(j)}(k) +  H^{(j)}(k+1).
\label{rec_rels_H_k}
\ee

The column vectors of $H(0)$ are 
\be
H^{(j)} (0)= \begin{pmatrix}h_0^{(j-1)} \cr \vdots \cr h^{(j-1)}_{-j+1} =1 \cr 0 \cr  \vdots\cr 0 \end{pmatrix} := h^{(j-1)}.
\ee
Therefore 
\be
H(0)=  \left( h^{(0)} \ h^{(1)} \dots h^{(n-1)}\right) 
\ee
is upper triangular with $1$'s along the diagonal and hence has unit determinant
\be
\det\left(H(0)\right) = 1.
\ee
Note that, because of (\ref{negative_h0}), the Pl\"ucker coordinates $\pi_\lambda(H)$ of the element $H$
may equivalently be written  as $\ell(\lambda) \times \ell(\lambda) $ determinants:
\be
\pi_{\lambda}(H) = \det (H_\lambda)  = \det \begin{pmatrix}h_{\lambda_1 }^{(0)} &h_ {\lambda_1 }^{(1)} & \cdots & h_{\lambda_1 }^{(\ell(\lambda)-1)}  \cr
\vdots & \vdots & \cdots & \vdots \cr 
h_{\lambda_{\ell(\lambda)} -\ell(\lambda)+1}^{(0)} & h_{\lambda_{\ell(\lambda)} -\ell(\lambda)+1}^{(1)} &  \cdots & 
h_{\lambda_{\ell(\lambda)} -\ell(\lambda)+1}^{(\ell(\lambda)-1)} 
\end{pmatrix}.
\ee

We are now ready to state the main result, which is a generalization of the one obtained
for the case of orthogonal polynomials in \cite{SV}. (Cf.~also \cite{AKLTZ} for the general
setting of generalized Jacobi-Trudi identities associated to MKP $\tau$-functions.)
\bp (Quantum Jacobi-Trudi identity)
\label{H_Phi}
The semi-infinite matrices $H$ and $\Phi$ represent the same element $[H]=[\Phi] \in  \Gr_n(\HH_+,\Fb)$
of the Grassmannian. Since both $\Phi(0)$ and $H(0)$ are invertible, this means
\be
H H(0)^{-1} = \Phi \Phi(0)^{-1}.
\label{HH0_PhiPhi_0}
\ee
It follows that their Plucker coordinates coincide:
\be
\det \left(H_\lambda\right) =  {\det \left(\Phi_\lambda\right) \over\det \left(\Phi_0 \right)} = S^\phi_\lambda,
\ee
which is equivalent to the generalized  Jacobi-Trudi identity
\be
S^\phi_\lambda = \det\left( h^{(j-1)}_{\lambda_i -i +1}\right)|_{1\le i, j \le \ell(\lambda)}. 
\label{QJT}
\ee
\ep 
  \noindent{\bf Proof:} Eq.  (\ref{HH0_PhiPhi_0}) is equivalent to the set of equations
  \be
  \Phi(k) \Phi(0)^{-1} H(0) = H(k), \quad  k=0, 1, \dots
  \label{main_reln_k}
  \ee
  or, by columns
  \be
  \Phi(k) \Phi(0)^{-1} H^{(j)}(0) = H^{(j)}(k), \quad k=0, 1, \dots, \quad 1 \le j \le n.
   \label{main_reln_jk}
  \ee
We prove this by (finite) induction on $j$, for $1\le j \le n$. The $j=1$ case
\be
\Phi(k) \Phi(0)^{-1} H^{(1)}(0) = H^{(1)}(k), \quad \forall k=0, 1, \dots
\ee
  is satisfied  since, by Cramer's rule, it is equivalent to  the definition (\ref{h0i})
  of $h^{(0)}_{i}$. 
  
  Now assume (\ref{main_reln_jk}) holds for $j-1$
   \be
   \Phi(k) \Phi(0)^{-1} H^{(j-1)}(0) = H^{(j-1)}(k), \quad  k=0, 1, \dots. 
    \ee
   From  the recursion relations (\ref{rec_rels_Phi_k}) and (\ref{rec_rels_H_k})
   for $k=0$, we have
   \be
   H^{(j)}(0) = H^{(j-1)}(1) - \Phi(1) \Phi^{-1}(0) H^{(j-1)}(0) + \Phi(0) X \Phi^{-1}(0) H^{(j-1)}(0).t
   \ee
   By the inductive hypothesis, the first two terms cancel, so we have
   \be
   H^{(j)}(0) = \Phi(0) X \Phi^{-1} (0) H^{(j-1)}(0).
   \ee
   Substituting this in the LHS of (\ref{main_reln_jk}) gives
   \bea
     \Phi(k) \Phi^{-1}(0) H^{(j)}(0) &\& = \Phi(k)X \Phi^{-1}(0) H^{(j-1)}(0) \cr
     &\& = \Phi(k+1) \Phi^{-1}(0) H^{(j-1)} (0) + J(k) \Phi(k) \Phi^{-1}(0) H^{(j-1)}  \cr
     &\& = H^{(j-1)}(k+1)+ J(k) H^{(j-1)}(k)\cr
     &\&  = H^{(j)}(k),
     \eea
     where the recursion relation (\ref{rec_rels_Phi_k}) was used in the second line,
     the inductive hypothesis in the third,  and the recursion relation (\ref{rec_rels_H_k})
     in the fourth. \hfill QED
     
     We now give an alternative proof of Proposition (\ref{H_Phi}),  which sheds further
     light on the meaning of the elements $[\Phi]=[H]$. 
     For the case of monomials,
 denote the matrix $H$ as $\overset{\scriptscriptstyle 0}{H}$.
We then have the following:
\bl
\label{HAPhi}
\be
H = A^\phi \overset{\scriptscriptstyle 0}{H}
\label{HH0}
\ee
or equivalently, column by column
\be
H^{(j)}= A^\phi \overset{{}_0}{H}{}^{(j)}, \quad j=1, \dots n.
\label{HjA_recurs}
\ee
\el
The proof is again inductive in the columns. For $j=1$,  by elementary row operations,
\bea
&\&\det \begin{pmatrix} \phi_{n+k}(x_1) & \phi_{n+k}(x_2)& \cdots &\phi_{n+k}(x_n) \cr
\phi_{n-2}(x_1) & \phi_{n-2}(x_1)& \cdots &\phi_{n-2}(x_n) \cr
\vdots & \vdots & \cdots & \vdots \cr
\phi_{1}(x_1) & \phi_{1}(x_1)& \cdots &\phi_{1}(x_n) \cr
1 & 1& \cdots &1 \cr
\end{pmatrix} \cr
&\& \phantom{P}\cr
=&\&\det \begin{pmatrix} \phi_{n+k}(x_1) & \phi_{n+k}(x_2)& \cdots &\phi_{n+k}(x_n) \cr
x_1^{n-2} &  x_2^{n-2}& \cdots &x_n^{n-2} \cr
\vdots & \vdots & \cdots & \vdots \cr
x_1 & x_2& \cdots &x_n \cr
1 & 1& \cdots &1 \cr
\end{pmatrix} \cr 
=&\&  \sum_{i=0}^{n+k} A^\phi_{n+k+1, i+1} h_i \Delta(x_1, \dots, x_n),
\eea
where
\be
h_i := {\det \begin{pmatrix} x_1^{n+i-1} & x_2^{n+i-1}& \cdots &x_n^{n+i-1} \cr
x_1^{n-2} &  x_2^{n-2}& \cdots &x_n^{n-2} \cr
\vdots & \vdots & \cdots & \vdots \cr
x_1 & x_2& \cdots &x_n \cr
1 & 1& \cdots &1 \cr
\end{pmatrix}\over \Delta(x_1, \dots x_n)}
\ee
are the complete symmetric functions, which are the components of $\overset{{}_0}{H}{}^{(0)}$.

Now assume relation (\ref{HjA_recurs}) holds for $j-1$,
\be
H^{(j-1)}= A^\phi \overset{{}_0}{H}{}^{(j-1)}.
\ee
Multiplying both sides on the left by $J$, and using (\ref{A_Phi_J_A}) gives
\be
H^{(j)}=JH^{(j-1)} = JA^\phi \overset{{}_0}{H}{}^{(j-1)} = A^\phi\Lambda^t \overset{{}_0}{H}{}^{(j-1)}  = A^\phi\overset{{}_0}{H}{}^{(j)} .
\ee
From this, it follows that
\be
H = \Phi T,
\ee
where
\be
T= \Phi(0)^{-1}H(0) = \overset{{}_0}{\Phi}(0){}^{-1}\overset{{}_0}{H}(0).
\ee

This shows, in particular, that the change of basis matrix $\Phi(0)^{-1}H(0)$ is
independent of the polynomial system $\{\phi_i(x)\}$ chosen.  The invertible
upper triangular matrix $A^\phi$ may be viewed as defining a group element
$G_\Phi \in GL(\HH_+, \Fb)$ that carries the bases $\overset{{}_0}{H}$ and $\overset{{}_0}{\Phi}$
into $H$ and $\Phi$, respectively. 
\br
Eq. (\ref{HH0}) could have been taken as the starting definition of $H$,
from which the recursion relations (\ref{rrHj}) follow as a consequence of the intertwining
relation  (\ref{A_Phi_J_A}).
\er

   In the context of integrable systems, the intertwining relation $(\ref{A_Phi_J_A})$ may
   be viewed as a ``dressing'' transformation, that produces the initial value of the Lax matrix 
    $J$ from  the ``bare'' (or vacuum) solution flowing from $\Lambda^t$. The dynamics are determined by
an abelian group action on the Grassmannian, which induces commuting flows of KP type,
as well as an isospectral flow of generalized Toda type for $J$ \cite{AvM}

\subsection{The generalized Jacobi-Trudi formula: dual form}
We now give a dual form of (\ref{QJT}) that can be written in terms of the conjugate partition and analogs of the \textit{elementary symmetric functions}.  Let $\overset{{}_0}{\tilde{E}}$  denote the $\infty \times \infty$ matrix
with row vectors denoted $\overset{{}_0}{E}_{(i)}$, $i \in \Zb,  -\infty < i \le n$, with
$i$ decreasing from $n$  to $-\infty$ vertically upward.
  \begin{equation}
  \overset{{}_0}{E}{}:=\left(
                                  \begin{array}{c}
                                  \vdots\\
                                      \overset{{}_0}{E}{}_{(n-1)} \\
                                       \overset{{}_0}{E}{}_{(n)} \\
                                      \end{array}
                                \right).
 \end{equation}
The components of  the row vectors $\overset{{}_0}E_{(i)}$ are defined to be
\be
 \overset{{}_0}{E}{}_{(i)}^{j} = (-1)^{n-i-j+1}e_{n-i-j+1},~~j\in \mathbb{N},~~ - \infty <i \leq n, 
 \label{E_0_def}
\ee
 where $e_{i}$ is the $i$th \textit{elementary symmetric function}. The ordering
 is such that the column index $j$ decreases from left to right,  ending with $j=1$  
 (i.e., the transpose of the convention for matrices $\overset{{}_0}{\bf H}$ and
 ${\bf H}$).  Note that the bottom row $\overset{{}_0}E_{(n)}$  is just the semi-infinite
 row vector
 \be
 \overset{{}_0}E_{(n)}  = (\dots, 0, 0, 1) = {\bf b}_1^t.
 \ee
 It follows from (\ref{E_0_def}) that the rows satisfy the recursion relations
 \bea
 \overset{{}_0}E_{(i)} &\&= \overset{{}_0}{E}{}_{(i+1)}\Lambda^t + \overset{{}_0}{E}{}_{(n)}(-1)^{n-i}e_{n-i} 
 \label{E_0_recurs_down}\\
 \overset{{}_0}{E}{}_{(i+1)} &\&= \overset{{}_0}{E}{}_{(i)}  \Lambda.
  \label{E_0_recurs_up}\
 \eea
 
 Now define the $\infty \times \infty$ matrix
 \be
  {\bf E} :=  \overset{{}_0}{\bf E}(A^\phi)^{-1}
  \label{E_def}
\ee
  whose rows are similarly denoted $E_{(i)}$, $i\in \Zb, \  i \le n$, and whose elements are
\be
  E_{(i)}^j= E_{ij}:=(-1)^{n-i-j+1}e_{(-i)}^{n-j+1}.
\ee
  Note that, because of the upper triangular form of $A^\phi$, we have
  \be
  E_{(n)} =  \overset{{}_0}E_{(n)}  = (\dots, 0, 0, 1).
  \ee
  \bp
 The rows of $E$ satisfy the recursion relations
  \bea
 E_{(i)} &\&=  E_{(i+1)} J + E_{(n)}(-1)^{n-i}e_{n-i} \\
 E_{(i+1)} &\&=  E_{(i)} \tilde{J} ,   \quad  -\infty < i\leq  n-1.
 \eea
 \ep
\noindent{\bf Proof:} From (\ref{E_def}), and (\ref{E_0_recurs_down}) we have
  \begin{eqnarray}
   E_{(i)} & = & \overset{{}_0}{E}{}_{(i)}(A^\phi)^{-1} ~=~  (\overset{{}_0}{E}{}_{(i+1)}\Lambda^t 
   + \overset{{}_0}{E}{}_{(n)}(-1)^{n-i}e_{n-i})(A^\phi)^{-1} \nonumber \\
    & =&E_{(i+1)} A^\phi\Lambda^t(A^\phi)^{-1} + E_{(n)}(-1)^{n-i}e_{n-i}  \\
    &=& E_{(i+1)} J + E_{(n)}(-1)^{n-i}e_{n-i}.
  \end{eqnarray}
From (\ref{E_def}), and (\ref{E_0_recurs_up}) we have
  \be
   E_{(i+1)} = \overset{{}_0}{E}{}_{(i+1)}(A^\phi)^{-1} = \overset{{}_0}{E}{}_{(i)}\Lambda (A^\phi)^{-1} =
   E_{(i)}A^\phi \Lambda^t (A^\phi)^{-1}
   =E_{(i)} \tilde{J}.
\ee

 \bc
The matrices ${\bf E}$ and ${\bf H}$ are mutual inverses
\be 
{\bf H}{\bf E} = {\bf E}{\bf H} = \Ib_\infty.
\label{HE_inverse}
\ee
\ec

\noindent{\bf Proof:} The complete and elementary symmetric functions
satisfy the orthogonality relations
\be
\sum_{k=i}^j (-1)^{i-k} e_{i-k} h_{k-j} = \delta_{ij}, 
\ee
which imply that $\overset{{}_0}{\bf H}  $ and $\overset{{}_0}{\bf E} $ are mutually inverse
\be
\overset{{}_0}{\bf H} \overset{{}_0}{\bf E} = \overset{{}_0}{\bf E} \overset{{}_0}{\bf H} = \Ib_\infty.
\ee
Eq.~(\ref{HE_inverse}) then follows from
\be
{\bf H}= A^\phi\overset{{}_0}{\bf H}, \quad {\bf E} = \overset{{}_0}{\bf E} (A^\phi)^{-1}.
\ee

The subspace   spanned by $\{ \overset{{}_0}{E}{}_{(0)}, \overset{{}_0}{E}{}_{(-1)},\cdots\}$ may be viewed 
as  the element $W^{*} \in \Gr^*_n(\HH_+,\Fb)$  of the dual Grassmannian that annihilates  the 
element $W\in \Gr_n(\HH_+,\Fb) $ spanned by 
$\{H^{(1)},\cdots,H^{(n)}\} $. It follows  \cite{GH} that the Pl\"{u}cker coordinates,  
$\pi_{\lambda}(W)$ and $\pi_{\lambda'}(W^{*})$ are related by 
 \begin{eqnarray}
\det\left( h^{(j-1)}_{\lambda_i -i +1}\right)|_{1\le i, j \le \ell(\lambda)} &=& \frac{\pi_{\lambda}(W)}{\pi_{0}(W)} = (-1)^{|\lambda|}\frac{\pi_{\lambda'}(W^{*})}{\pi_{0}(W^{*})} \nonumber \\
&=& (-1)^{|\lambda|} \det \Big((-1)^{\lambda'_j - j +i} e^{\lambda'_j-j+1}_{(i-1)}\Big)|_{1\leq i,j \leq l(\lambda')} \nonumber\\
 &=&\det \Big( e^{\lambda'_j-j+1}_{(i-1)}\Big)|_{1\leq i,j \leq l(\lambda')},
   \label{QJT_dual}
 \end{eqnarray}
 which yields the dual form of the generalized Jacobi-Trudi identity,.
 \bp (Dual generalized Jacobi-Trudi identity)
\label{E_Phi}
\be
S^\phi_\lambda = \det \Big( e^{\lambda'_j-j+1}_{(i-1)}\Big)|_{1\leq i,j \leq l(\lambda')} .
\label{dual_q_jacobi_trudi}
\ee
\ep 

\bc
The elements of the row $E_{(0)}$ are equal to the generalized elementary symmetric functions
$ S_{(1)^{n-j+1}}^\phi$
\be
E_{(0)}^{j} = e^{n-j+1}_{(0)} = S_{(1)^{n-j+1}}^\phi, \quad 1 \le j \le n .
\ee
\ec
  \noindent{\bf Proof:} This follows from (\ref{dual_q_jacobi_trudi}) by choosing the partition $\lambda = (1)^{n-j+1}$,
  and noting the dual partition $\lambda' = (j)$  has length $1$.

\section{Integrable systems and  $\tau$-functions}

\subsection{The polynomials $S^\phi_\lambda$ as KP $\tau$-functions }

We  now show that the generalized Schur polynomials $S^\phi_\lambda([x])$,
 when viewed as functions of the  monomial sums
\be
t_i := {1\over i}\sum_{a=1}^n x_a^i,
\label{monom_sums}
\ee
 are KP $\tau$-functions, in the sense of Sato  \cite{Sa, SS} and Segal and Wilson  \cite{SW}.  We use the standard notation
\be
[x] = {\bf t} = (t_1, t_2, \dots )
\label{monom_sums_x}
\ee
 to denote the infinite sequence of monomials sums defined in (\ref{monom_sums}).  First, we define coefficients $\phi^{(n)}_{\lambda \mu}$
as the Pl\"ucker coordinates $\pi_\mu(C^{(\lambda, n)})$ of the element $C^{(\lambda,n)}$ of  the Grassmannian $\Gr_n(\HH_+,\Cb)$
 spanned by the polynomials $\{\phi_{\lambda_i -i+n}(z)\}_{i=1, \dots n}$
\be
\phi^{(n)}_{\lambda \mu} := \pi_\mu(C^{(\lambda, n)}) = \det \left( \phi_{l_i, m_j} \right)|_{1\le i,j \le n} ,
\ee 
where
\be
l_i := \lambda_i - i +n,   \quad  m_j := \mu_j - j +n,  \quad i, j \in \Nb^+, 
\ee
are the particle coordinates associated to partitions $\lambda$ and $\mu$.

So far, we have not explicitly indicated the dependence on the integer $n$ in the expression
for the symmetric polynomials $S^\phi_\lambda([x])$. But henceforth, $n$ will
be viewed as an integer variable, and we  use the following notation to indicate this
\be
S^\phi_{\lambda, n}([x]) := S_\lambda^\phi (x_1, \dots x_n).
\ee
\br
Schur polynomials $S_\lambda([x])$, when viewed as
functions of the monomial sum variables (\ref{monom_sums}), (\ref{monom_sums_x}),
do not depend on the number $n$ of $x_i$'s, provided this is at least equal to 
the length $\ell(\lambda)$ of the partition. (Otherwise, they vanish). However, for general 
$S^\phi_{\lambda,n}([x])$, this is not the case, even though these too may be viewed
as functions of the monomial sum variables (\ref{monom_sums}), (\ref{monom_sums_x}).
As will be seen explicitly below, the $S^\phi_{\lambda,n}([x])$'s do, in general, depend on the number $n$. 
(The $n$ independence in the case of $\Phi=\overset{{}_0}{\Phi}$ is due to the fact that, 
in this case, we have the equality
\be
h^{j}_i = h^{(0)}_{j+i} = h_{j+i},
\ee
where the $h_i$'s are the usual complete symmetric functions.)
\er
The following shows that the $\phi^{(n)}_{\lambda \mu}$'s are the coefficients  in the expression 
for  $S_{\lambda, n}^\phi([x])$ as a linear combination of  Schur functions $S_\lambda([x])$.
\bl
\label{Sphi_Schur}
\be
S_{\lambda,n}^\phi([x]) =\sum_{\mu \atop {\ell(\mu)\le n \atop|\mu| \le |\lambda |}} \phi^{(n)}_{\lambda \mu} S_\mu([x]).
\label{Schur_exp_Sphi}
\ee

\el
\noindent{\bf Proof:} Let $C^{(\lambda,n)}$ be
the $\infty \times n$ matrix whose $i$th column  consists of the coefficients $\phi_{\lambda_i -i +n +1, j}$ 
of the polynomial $\phi_{\lambda_i-i + n +1}(z) $ (i.e.,  its representation as a column vector relative
to the basis $\{{\bf b}_j :=  z^{j-1}\}_{j\in \Nb}$).
By the Cauchy-Binet identity, we have
\be
\det\left((C^{(\lambda,n)})^t \overset{{}_0}{\Phi} \right)=
  \sum_{ {\mu} \atop \ell(\mu) \le n} \det\left((C^{(\lambda,n)})_\mu\right)\  \det  (\overset{{}_0}{\Phi}_\mu)
=\sum_{\mu \atop {\ell(\mu)\le n \atop|\mu| \le |\lambda |}}\pi_\mu(C^{(\lambda,n)}) S_\mu([x]). 
\ee
\bc
$S_{\lambda, n}^\phi({\bf t})$  is a KP $\tau$-function.
\ec
This follows from the fact that any expansion of the form (\ref{Schur_exp_Sphi}) is a KP $\tau$-function provided
the coefficients $\phi_{\lambda \mu}$ satisfy the Pl\"ucker relations (with respect to $\mu$, for fixed $\lambda$). 
\br
Note that, unlike ordinary Schur polynomials $S_\lambda([x])$, the polynomials
$S^\phi_\lambda([x]) $ do not necessarily vanish when $[x]=[{\bf 0}]$ for $\lambda\neq (0)$.
By Lemma \ref{Sphi_Schur}, we have

\be
S^\phi_{\lambda}({\bf 0}) = \phi^{(n)}_{\lambda, (0)} =\det\left(\phi_{\lambda_i -i +n, n-j}\right)|_{1\le i,j \le n}.
\ee
\er
\br
 Through a subquotienting procedure \cite{BFH}, it is possible to identify the  elements of the infinite 
 Grassmannian $\Gr_{\HH_+}(\HH_- +\HH_+)$ of Sato  \cite{Sa, SS}and Segal-Wilson \cite{SW} corresponding 
 to these polynomial $\tau$-functions. Equivalently, they can be expressed as  vacuum state expectation values 
 of products of fermionic operators, as explained in the next section.
\er

\subsection{Fermionic representation of $\tau$-functions }

The coefficients $ \phi^{(n)}_{\lambda \mu}$ can  be expressed in terms of fermionic creation and annihilation operators 
$\{\psi_i, \psi^\dag_i\}_{i \in \Zb}$ on a Fermi Fock space, satisfying the usual  anticommutation relations
\be
[\psi_i, \psi^\dag_j]_+ = \delta_{ij}, \quad [\psi_i, \psi_j]_+ = 0,  \quad [\psi^\dag_i, \psi^\dag_j]_+ = 0, \quad i,j \in \Zb, 
\label{canon_anticom}
\ee
with the vacuum state $|0\rangle$ satisfying
\be
\psi_{i} | 0 \rangle = 0, \quad i <0, \quad \psi^\dag_i |0\rangle, \quad i\ge 0.
\ee
For $n>0$, let
\be
|n\rangle :=  \psi_{n-1} \cdots \psi_{0} |0\rangle, \quad   |-n\rangle :=  \psi^\dag_{-n} \cdots \psi^\dag_{-1} |0\rangle
\ee
be the charge $n$  (or $-n$) vacuum state, and denote the basis states in the charge $n$ sector (for $n$ positive
or negative)
\be
|\lambda;n\rangle := \psi_{\lambda_1-1 +n} \cdots \psi_{\lambda_{\ell(\lambda)}-\ell(\lambda)+n} |n-\ell(\lambda)\rangle.
\label{lambda_n_ket}
\ee
(In particular, $|(0);n\rangle$ is just denoted $|n\rangle$.)
Recall that in general a lattice of mKP $\tau$-functions can be expressed \cite{DJKM, JM} as 
\be
\tau_{g}(n,{\bf t}) = \langle n | \gamma_+ ({\bf t}) g | n\rangle, \quad n\in \Zb,
\ee
where 
\be
\gamma_+({\bf t}) :=e^{\sum_{i=1}^\infty t_i J_{i}} , \quad
J_i  := \sum_{j\in \Zb} \psi_j \psi_{j+i}^\dag
\ee
and $g$ is any element of the infinite Clifford algebra generated  by (\ref{canon_anticom})
that satisfies the bilinear relation
\be
\left[ \sum _{i\in \Zb} \psi_i \otimes \psi^{\dag}_i , g \otimes g \right] = 0.
\ee
In particular, the Schur functions can be expressed fermionically either as
\be
 S_\lambda({\bf t})= \langle n | \gamma_+({\bf t })|\lambda; n\rangle, 
\label{Schur_fermion}
\ee
or equivalently as
\be
 S_\lambda({\bf t})=  \langle \lambda;n | \gamma_-({\bf t}) |n\rangle,
\label{Schur_fermion}
\ee
where
\be
\gamma_-({\bf t}) := e^{\sum_{i=1}^\infty t_i J_{-i}}.
\label{J_i_def}
\ee
A 2D-Toda chain of $\tau$-functions can be represented fermionically \cite{DJKM} as 
\be
\tau_{g}(n,{\bf t}, {\bf s}) = \langle n |  \gamma_+({\bf t}) g\gamma_- ({\bf s}) | n\rangle, \quad n\in \Zb. 
\ee
where
\be
{\bf s} := (s_1, s_2, \dots )
\ee
is  viewed as an additional infinite set of commuting flow variables.

By Wick's theorem, we have (cf. ref. \cite{BFH})
\be
\phi^{(n)}_{\lambda \mu} = \langle \mu; n | \prod_{i=1}^n w^\lambda_i | 0 \rangle,
\label{phi_n_lambda_mu}
\ee
where
\be 
w^\lambda_i := \sum_{j=0}^{\lambda_i -i +n}\phi_{\lambda_i -i +n, j} \psi_j.
\label{w_mu_i}
\ee

We may  also express $\phi^{(n)}_{\lambda \mu}$ in  a more symmetrical form as follows.
The matrix $\phi$  may be viewed as representing an element of the group of  invertible automorphisms of the
Hilbert space $\HH:= L^2(S^1) $  of square integrable functions
on the unit circle $\{z | \  |z|^2 =1\}$, which is the direct sum of the two subspaces
$\HH_+$ and $\HH_-$ spanned, respectively, by the positive powers $ \{z^i\}_{i\in \Nb}$
and negative powers $\{z^{-i}\}_{i \in \Nb^+}$ of $z$ . The element represented by
$\phi$ in this monomial basis, leaves each subspace $\HH_{\pm}$ invariant, acts
 as the identity  transformation within the subspace $\HH_+$  and has matrix representation
$\phi$ in the negative monomial basis $\{z^{-i}\}$ for $\HH_-$.
We may express this infinite  lower triangular  matrix with coefficients $(\phi_{ij})_{i,j \in \Nb}$ 
that are all equal to $1$ on the diagonal as the exponential of a strictly lower triangular matrix  $\alpha$
\be
\phi = e^\alpha
\ee
 with matrix elements $(\alpha_{ij})_{i,j \in \Nb}$ satisfying
\be 
\alpha_{ij} = 0 \quad {\rm if }\quad  j\ge i.
\ee
The fermionic representation  \cite{ DJKM, HO1, Sa} of this group element  is then
\be
g_\phi:=\exp{\sum_{i >j\ge 0}\alpha_{ij} \psi_i\psi^\dag_j}.
\label{g_phi}
 \ee
 while the fermionic representation of the transpose
 \be
\phi^t =  e^{\alpha^t}
\label{g_phi_t}
\ee
is 
\be
g_{\phi^t} = \exp{\sum_{i >j\ge 0}\alpha_{ij} \psi_j \psi^\dag_i}.
\ee
Note that $g_\phi$ stabilizes the left charge $n$ vacuum and $g_{\phi^t}$ the right charge $n$ vacuum for any $n$
\be
 \langle n | g_\phi = \langle n|, \quad  g_{\phi^t}  |n\rangle = |n \rangle
 \label{g_phi_stab_vac}
\ee
since, for $i>j\ge 0$
\be
\psi_j \psi^\dag_i |n\rangle = 0.
\ee

\bl
The matrix  elements $A^\phi_{ij}$ and $J_{ij}$ may be expressed fermionically as 
\bea
A^\phi_{ij} &\&=\phi_{i-1, j-1} = \langle (i); 0| g_\phi | (j); 0\rangle = \langle (i-k); k| g_\phi | (j-k ); k\rangle,
 \label{Aij_fermion} \\
 J_{ij} &\& = \langle  (i )  ; 0 | g_\phi J_{-1} g_\phi^{-1}| (j) ; 0\rangle=
  \langle  (i - k)  ; k | g_\phi J_{-1} g_\phi^{-1}| (j-k) ; k\rangle,  \quad \forall \ k\le i, j, 
 \label{Jij_fermion}
\eea
\el
for any charge  sector $k\le i, j$.

\noindent {\bf Proof:}
Eq.~(\ref{Aij_fermion}) follows from eqs.~(\ref{lambda_n_ket}), (\ref{g_phi_stab_vac}) and (\ref{g_phi}),    which imply
 \bea
 \langle (i-k); k| g_\phi| (j-k);k\rangle  
&\& =  \langle (0); k-1| \psi_{i-1}^\dag g_\phi \psi_{j-1} | (0);k -1\rangle \cr
 &\&  =   \langle (0); k-1|g^{-1}_\phi  \psi_{i-1}^\dag g_\phi \psi_{j-1} | (0);k -1\rangle \cr
 &\& =   \sum_{l =0}^{i-1} \phi_{i-1, l} \langle (0); k-1| \psi^\dag_l  \psi_{j-1} | (0);k -1\rangle. \cr
 \&\& = \phi_{i-1, j-1} = A^\phi_{ij}.
 \eea
Eq.~(\ref{Jij_fermion})  follows by substituting
\be
g_\phi J_{-1}g_\phi^{-1} = \sum_{l\in \Zb} g_\phi \psi_l g_\phi^{-1} g_\phi \psi^\dag_{l-1} g_\phi^{-1}   = \sum_{k=0}^\infty \sum_{l=0}^{k} \sum_{m=0}^{l-1} \phi_{kl} \phi^{-1}_{l-1, m} \psi_k \psi^\dag_m
\ee
into 
 \bea
 \langle  (i -k)  ; k | g_\phi J_{-1} g_\phi^{-1}| (j -k ) ; k\rangle 
 &\& = \sum_{k=0}^\infty \sum_{l=0}^{k} \sum_{m=0}^{l-1}  \phi_{kl} \phi^{-1}_{l-1, m} \langle  (i -k )  ; k |  \psi_k \psi^\dag_m    | (j-k) ; k\rangle  \cr
&\&= \sum_{k=0}^\infty \sum_{l=0}^{k} \sum_{m=0}^{l-1}  \phi_{kl} \phi^{-1}_{l-1, m} \langle (0); k-1| \psi_{i-1}^\dag   \psi_k \psi^\dag_m    \psi_{j-1} | (0);k -1\rangle  \cr
&\&= \sum_{k=0}^\infty  \sum_{m=0}^{l-1}  \phi_{i-1,l} \phi^{-1}_{l-1, j-1} 
= \left(A^\phi \Lambda^t (A^\phi)^{-1}\right)_{ij} = J_{ij},
\eea
where the third equality follows from Wick's theorem.

More generally, we have the following expression for $\phi^{(n)}_{\lambda \mu}$ as a fermionic matrix element
\bp
\label{g_phi_lambda_mu_prop}
\be
\phi^{(n)}_{\lambda \mu} = \langle \lambda;n |g_\phi | \mu;n\rangle  =\langle \mu;n | g_{\phi^t} | \lambda; n \rangle
\label{g_phi_lambda_mu}.
\ee

\ep
\noindent{\bf Proof:}{}\footnote{This proof was suggested by A. Yu. Orlov.  See also \cite{HO2}, eq. (1.34) 
for a related fermionic identity.}

It follows from the canonical anticommutation relations (\ref{canon_anticom}), that
\be
g_{\phi^t} \psi_i g^{-1}_{\phi^t} = \sum_{j=0}^i \phi_{ij} \psi_j . 
\label{dressing_psi_i}
\ee
Therefore, from eq. (\ref{w_mu_i}), we have
\be
w^\lambda_i := 
 \sum_{j=0}^{\lambda_i-i +n} \phi_{\lambda_i -i +n, j} \psi_j 
 =g_{\phi^t}\psi_{\lambda_i -i +n} g_{\phi^t}^{-1}.
\ee
The expression (\ref{lambda_n_ket}) for $|\lambda;n\rangle$ can equivalently be written
\be
|\lambda;n\rangle := \psi_{\lambda_1-1 +n} \cdots \psi_{\lambda_n} |0\rangle.
\ee
Therefore, from (\ref{phi_n_lambda_mu}), 
\be
\phi^{(n)}_{\lambda \mu} = \langle \mu; n | w_1^\lambda \dots w_n^\lambda |n\rangle
= \langle \mu;n | g_{\phi^t} | \lambda;n\rangle  = \langle \lambda;n| g_\phi | \mu; n\rangle.
\ee

\hfill QED

From (\ref{Schur_fermion}) and (\ref{g_phi_lambda_mu}),  we  obtain a fermionic expression 
for $S^\phi_{\lambda,n}([x])$.
\bc
\be
S^\phi_{\lambda, n}([x]) = \langle \lambda;n | g_\phi \gamma_-([x]) | n \rangle 
= \langle  n |  \gamma_+([x] ) g_{\phi^t}  |\lambda; n\rangle  .
\label{S_phi_n_fermionic}
\ee
\ec

Moreover, the polynomials $S^\phi_\lambda([x])$  can themselves be used as coefficients 
in a Schur function expansion to define a  family of KP $\tau$-functions, 
in which the indeterminates $(x_1, \dots x_n)$ are interpreted as complex parameters
\be
\tau_\phi(n,  {\bf t}, [x]) := \sum_{\lambda} S_{\lambda, n}^\phi([x]) S_\lambda({\bf t}).
\label{s_tau_phi}
\ee
Here, the KP flow parameters ${\bf t} = (t_1, t_2, \dots)$ are independent and 
\be
{\bf s} := (s_1, s_2, \dots ) := [x]
\ee
may be viewed as a second set of flow parameters. Then $\tau_\phi(n, {\bf t}, {\bf s}) $ is simultaneously
a KP $\tau$-function in both  ${\bf s}$ and ${\bf t}$ variables and, viewing  $n$ as a
lattice variable, a 2D Toda $\tau$-function.
\bp
The functions $\tau_\phi(n, {\bf t}, {\bf s}) $  form  a 2D-Toda chain of $\tau$-functions
which may be expressed fermionically as
\be
\tau_\phi(n, {\bf t}, {\bf s}) =\langle n|\gamma_+({\bf t}) g_\phi \gamma_-({\bf s}) |n \rangle.
\label{tau_phi_Schur_exp}
\ee
\ep
\noindent {\bf Proof:}
We substitute (\ref{Schur_exp_Sphi}) in (\ref{s_tau_phi}) to express
$\tau_\phi(n, {\bf t}, {\bf s})$ as a double Schur function expansion
\be
\tau_\phi(n, {\bf t}, {\bf s}) = \sum_{\lambda} \sum_\mu \phi^{(n)}_{\lambda \mu}S_\lambda({\bf t}) S_\mu({\bf s}) .
\ee
Using eq.~(\ref{g_phi_lambda_mu})  and (\ref{Schur_fermion}) gives
\bea
\tau_\phi(n, {\bf t}, {\bf s}) &\& =\sum_{\lambda} \sum_\mu \langle n | \gamma_+({\bf t })     
 |\lambda ;n \rangle   \langle \lambda;n | g_\phi | \mu;n \rangle \langle \mu;n | \gamma_-({\bf s}) |n \rangle \cr
 &\& = \langle n|\gamma_+({\bf t}) g_\phi \gamma_-({\bf s}) |n \rangle,
\eea
which is the standard fermionic form \cite{DJKM, HO2} for  a chain of 2D-Toda $\tau$-functions.
\hfill QED

If all the parameters ${\bf s}$ in (\ref{tau_phi_Schur_exp}) are set equal to zero, we obtain the
chain of KP $\tau$-functions
\be
\tau_\phi(n, {\bf t}, {\bf 0}) =\langle n|\gamma_+({\bf t}) g_\phi |n \rangle
= \sum_{\lambda \atop \ell(\lambda)\le n} \phi^{(n)}_{\lambda, (0)}S_\lambda({\bf t}).
\label{tau_phi_0}
\ee

  Finally, we may choose a pair of polynomial systems  $\{\phi_i\}_{i\in \Nb}$ and  $\{\theta_i\}_{i\in \Nb}$ 
  and associate to them the generalized Schur functions $S^\phi_{\lambda,n}$ and $S^\theta_{\lambda,n}$.
  Forming the sum of their products
  \be
  \tau_{\phi, \theta}(n, {\bf t } ,{\bf s}) := \sum_\lambda S^\phi_{\lambda,n}({\bf s})S^\theta_{\lambda,n}({\bf t}),
  \ee
  we obtain a more general form of 2D Toda $\tau$-functions.
  \bp
  \be
  \tau_{\phi, \theta}(n, {\bf t }, {\bf s}) = \langle n | \gamma_+({\bf t}) g_\theta^t g_\phi \gamma_-({\bf s}) | n \rangle.
  \label{tau_theta_phi}
  \ee
  \ep
  \noindent{\bf Proof:} Substitute eq.(\ref{S_phi_n_fermionic}) in its first form for $S^\phi_{\lambda,n}({\bf s})$
  and in its second form for  $S^\theta_{\lambda,n}({\bf t})$ and use
  \be 
  \sum_\lambda | \lambda; n\rangle \langle \lambda; n | = \Pi_{(n)}
  \ee
where  $\Pi_{(n)}$ is the orthogonal projection map to the charge $n$ sector.

\br
The product $g_\theta^t g_\phi $ appearing in (\ref{tau_theta_phi})
represents a lower/upper triangular factorization
of an operator in which the diagonal elements  
in each term have coefficients $1$. This restriction, which follows from choosing the polynomials systems
$\phi$ and $\theta$  to be monic,  can be dropped  with no essential change in the resulting $\tau$-functions, 
except for their normalization,  since these  represent the same elements of the Grassmannian.
This amounts to mutiplying one of the factors $g_\theta^t $ or $g_\phi $, either on the left
or on the right,  by a diagonal term  of the form
\be
g_0(\rho) := e^{\sum_{i=0}^\infty T_i \psi_i \psi_i^\dag },
\ee
where the coefficients $\{T_i\}_{i\in \Nb}$ determine the leading term coefficients of the polynomials
\be
\phi_i(x) = e^{T_i} x^i + \OO(x^{i-1}) .
\ee
This  represents a convolution product  on $\HH= L^2(S^1)$ with the (distributional)  
element  $\rho$  whose negative Fourier series part 
 is $\rho_-(z) := \sum_{i=0}^\infty\rho_i z^{-i-1}$, and whose  positive part
is  the distribution represented by $\rho_+(z) = \sum_{i=0}^\infty  z^i$.
 Such convolution actions have been studied as symmetries of 
KP  and 2D Toda $\tau$-functions in \cite{HO3}.
The factor $g_0(\rho)$  may be placed on the left  $g_0(\rho)g_\theta^t g_\phi $ 
or on the right  $g_\theta^t g_\phi  g_0(\rho)$ in (\ref{tau_theta_phi})
without altering the symmetric polynomials $S^\theta_{\lambda,n}$ or $S^\phi_{\lambda,n}$
except by normalization factors. If the diagonal factor is placed in the
middle $g_\theta^t g_0(\rho)g_\phi $, this amounts to replacing one of
the two series of polynomials $\phi$ or $\theta$ by their convolution product
with $\rho$.
\er
\subsection{Fermionic representation of the matrices $H$ and $E$}

We now give fermionic representations of the matrices $H$ and $E$
appearing in the generalized Jacobi-Trudi formulae (\ref{QJT}), (\ref{QJT_dual}).
\bp
\label{EH_fermion}
The matrix elements $H^{(j)}_i$ and $E^j_{(i)}$ are given as fermionic matrix
elements as follows
\bea
H^{(j)}_i &\&= \langle (i+j-n-1) ; n-j +1 | g_\phi \gamma_-([x]) | n-j +1\rangle = h^{(j-1)}_{i-n}  \\
\label{Hij_fermion}
E^{j}_{(i)} &\&=(-1)^{n-i-j+1} \langle (1)^{n- i-j+1}; n-i | g_\phi \gamma_-([x]) | n-i \rangle   =(-1)^{n-i-j+1}e_{(-i)}^{n-j+1}.
\label{Eij_fermion}
\eea
\ep
\noindent{\bf Proof:}  We first verify these expressions for the case of monomials;
i.e. when $\Phi  = \overset{{}_0}{\Phi}$, and then show that eq.~(\ref{HjA_recurs}) of Lemma
\ref{HAPhi} is satisfied.

Setting $g_\phi= \Ib$ in (\ref{Hij_fermion}) we get
\be
 \langle (i+j-n-1) ; n-j +1 |\gamma_-([x]) | n-j +1\rangle
 = h_{i+j-n-1} ([x]) = \overset{{}_0}{H}{}^{(j)}_i.
\ee
From eq.~({\ref{lambda_n_ket}) it follows that
\bea
g_{\phi^t} | (i+j-n-1) ; n-j +1 \rangle &\&=g_{\phi^t} \psi_{i-1} |n- j\rangle \cr
&\&= \sum_{k=1}^{i} \phi_{i-1, k-1} |(k+j-n -1); n-j +1\rangle,
\eea
where (\ref{dressing_psi_i}}) has been used in the second line.
Therefore
\bea
\langle (i+j-n -1) ; n-j +1 | g_\phi \gamma_-([x]) | n-j +1\rangle
&\&= \sum_{k=1}^{i} A^\phi_{i k}\langle (k+j -n-1); n-j +1 | \gamma_-([x]) | n-j+1\rangle \cr
&\& = \sum_{k=1}^{i} A^\phi_{i k} \overset{{}_0}{H}{}^{(j)}_k = H^{(j)}_k .
\eea

To prove eq.~(\ref{Eij_fermion}) we proceed similarly. Setting $g_\phi= \Ib$ in (\ref{Eij_fermion}) we get
\be
(-1)^{n-i-j+1} \langle (1)^{n- i-j+1}; n-i | \gamma_-([x]) | n-i \rangle  = 
(-1)^{n-i-j+1}  e_{n-i-j+1} =  \overset{{}_0}{E}^{j}_{(i)}.
\ee
From eq.~({\ref{lambda_n_ket}) it follows that
\bea
g_{\phi^t}  |(1)^{n- i-j+1}; n-i  \rangle   &\& = - g_{\phi^t} \psi^\dag_{j-1} |( n-i +1\rangle  \cr
&\&= - \sum_{k=1}^{i} \phi^{-1}_{k-1, i-1} |(n-k-j+1); n-j+1\rangle,
\label{g_phi_enij}
\eea
where 
\be
g_{\phi^t} \psi^{\dag}_{j-1} g^{-1}_{\phi^t}= \sum_{k=0}^{\infty} \psi_{k-1}^\dag \phi^{-1}_{k-1, j-1}
\label{dressing_psi_i_dag}
\ee
has been used in the second line. Substituting (\ref{g_phi_enij}}) in (\ref{Eij_fermion}) gives
\bea
&\&(-1)^{n-i-j+1} \langle (1)^{n- i-j+1}; n-i | g_\phi \gamma_-([x]) | n-i \rangle \cr
&\&= \sum_{k=0}^{\infty}  \phi^{-1}_{k-1, j-1} (-1)^{{n-i -k+1}} \langle (1)^{n- i-k+1}; n-i |  \gamma_-([x]) | n-i \rangle\cr
&\& =  \sum_{k=0}^{\infty}    \overset{{}_0}{E}{}^{k}_{(i)} A^{-1}_{k j}  = E^j_{(i)}.
\eea
\hfill QED

\br
Proposition  \ref{EH_fermion}  shows that the generalized Jacobi-Trudi formulae (\ref{QJT}), (\ref{QJT_dual})
are special cases of those studied in \cite{AKLTZ}, with the group element
denoted $G$ in formula (3.16) of \cite{AKLTZ} given by
\be
G= g_\phi \gamma_-([x]).
\ee
\er

\section{Examples and applications}

We close with some  examples and applications of the above results.

\subsection{Character expansions of classical groups.}
As pointed out in \cite{SV}, irreducible characters of the classical groups
$O(2n)$, $O(2n+1)$ and $Sp(2n)$ can be viewed as examples of the generalized
Schur functions studied here, corresponding to certain systems of orthogonal polynomials.
In each case, we may view the parameters $(x_1, \dots, x_n)$ as coordinates  on
the maximal torus of a compact real form of the relevant group. The subgroup reductions 
\be
Sp(n) \supset U(n), \quad O(2n) \supset U(n), \quad O(2n+1) \supset U(n)
\ee
correspond to identifying the maximal torus of $Sp(n), O(2n)$ and $O(2n+1)$
with subgroups of the maximal torus of $U(2n)$ or $U(2n+1)$ consisting
of elements of the form $\diag(x_1, \dots, x_n, x_1^{-1}, \dots x_n^{-1})$
 or   $\diag(x_1, \dots, x_n, x_1^{-1}, \dots x_n^{-1}, 1)$.

The systems of orthogonal polynomials $ \{\phi_i^{Sp(2n)}(z)\}$, $ \{\phi_i^{SO(2n)}(z)\}$ and
$ \{\phi_i^{SO(2n+1)}(z)\}$ are expressed in terms of the variables
\be
z:= x + x^{-1}
\ee
as follows:
\bea
 \phi_i^{Sp(2n)}(z) &\&=\sum_{j=0}^{i} x^{i-2j} = \sum_{j=0}^i\phi_{ij}^{Sp(2n)}z^j, \cr
 \phi_i^{SO(2n)}(z) &\&=x^i + x^{-i} =  \sum_{j=0}^i\phi_{ij}^{SO(2n)}z^j , \cr
 \phi_i^{SO(2n+1)}(z) &\&=\sum_{j=0}^{2i} x^{i-j}=  \sum_{j=0}^i\phi_{ij}^{SO(2n+1)}z^j ,
  \eea
  where
  \bea
  \phi_{2i, 2j}^{Sp(2n)} &\&=  (-1)^{i+j} \left( i+j \atop  i-j \right),
  \quad \phi_{2i+1, 2j+1}^{Sp(2n)} =  (-1)^{i+j } \left( i+j +1 \atop  i-j \right), \cr
     &\& \cr
  \phi_{2i, 2j+1}^{Sp(2n)}  &\& =\phi_{2i+1, 2j}^{Sp(2n)} =0,
  \cr
     &\& \cr
  \phi_{2i, 2j}^{SO(2n)} &\& =   (-1)^{i+j}  {i \over j} \left( i+j-1 \atop  i-j \right) \ {\rm if } \  j>0, \quad
   \phi_{2i+1, 2j+1}^{SO(2n)} =   (-1)^{i+j}  {2i+1 \over 2 j+1} \left( i+j \atop  i-j \right),
    \cr
    &\& \cr
  \phi_{0, 0}^{SO(2n)} &\&=1, \quad  \phi_{2i, 0}^{SO(2n)} = 2 (-1)^i  \ \  {\rm if } \  i>0, \quad  \phi_{2i, 2j-1}^{SO(2n)}   =\phi_{2i-1, 2j}^{SO(2n)} =0, 
  \cr
      &\& \cr
   \phi_{2i, 2j}^{SO(2n+1)} &\&=  (-1)^{i+j} \left( i+j \atop  i-j \right),
  \quad \phi_{2i+1, 2j+1}^{SO(2n+1)} =  (-1)^{i+j } \left( i+j +1 \atop  i-j \right), \cr
     &\& \cr
   \phi_{2i, 2j+1}^{SO(2n+1)} &\&=  (-1)^{i+j+1} \left( i+j \atop  i-j -1\right),
  \quad \phi_{2i+1, 2j}^{SO(2n+1)} =  (-1)^{i+j } \left( i+j \atop  i-j \right).
  \eea
  
  The recursion matrices $J^{Sp(2n)}$, $J^{SO(2n)}$ and $J^{SO(2n+1)}$ for these orthogonal polynomials
  are given by
  \bea
  J_{ij} ^{Sp(2n)} &\& = \delta_{i+1, j} + \delta_{i,  j+1}, \cr
   J_{ij} ^{SO(2n)} &\& = \delta_{i+1, j} + \delta_{i,  j+1} +\delta_{i 2}\delta_{j 1},\cr
    J_{ij} ^{SO(2n+1)} &\& = \delta_{i+1, j} + \delta_{i,  j+1} - \delta_{i 1}\delta_{j 1}.
      \eea
  As noted in \cite{SV}, the corresponding generalized Schur functions coincide with the irreducible characters
\cite{L,  FH}:
  \bea
S_\lambda^{Sp(2n)}(z_1, \dots , z_n) &\&=\chi_\lambda^{Sp(2n)} (x_1, \dots , x_n) \\
S_\lambda^{SO(2n)}(z_1, \dots , z_n) &\&= \chi_\lambda^{SO(2n)} (x_1, \dots , x_n) \\
S_\lambda^{SO(2n+1)}(z_1, \dots , z_n)&\&= \chi_\lambda^{SO(2n+1)} (x_1, \dots , x_n) 
\eea
where 
\be
z_i := x_i + x_i^{-1}. \quad i=1, \dots ,n
\ee
and 
\be
S_\lambda^{G }(z_1, \dots , z_n) := S_\lambda^{\phi^G }(z_1, \dots , z_n)
\ee
for $G = Sp(2n), \ SO(2n), \  {\rm or} \  SO(2n+1)$. 
The generalised Jacobi-Trudi formula for these cases are equivalent,
   by elementary row operations, to the  determintal formulae for these characters given in  Props. 24.22, 24.33, 24.44
   of \cite{FH}.
   
    For any pair of partitions $(\lambda, \mu)$, and $G = Sp(2n), \ SO(2n), \  {\rm or} \  SO(2n+1)$, define
  \be
  \phi^G_{\lambda \mu} := \det\left( \phi^G_{\lambda_i -i +n, \mu_j -j +n }\right)_{1\le i,j, \le n}. 
  \ee
  
By Lemma (\ref{Sphi_Schur}) we have the expansions
\bea
S_\lambda^{Sp(2n)}(z_1, \dots , z_n) &\&= 
\sum_{ \ell(\mu)\le n \atop \mu\le |\lambda|} \phi^{Sp(2n)}_{\lambda \mu}
S_\mu(z_1, \dots,z_n), 
 \\
S_\lambda^{SO(2n)}(z_1, \dots , z_n)  &\&= 
\sum_{\ell(\mu)\le n \atop \mu\le |\lambda|}\phi^{SO(2n)}_{\lambda \mu}
S_\mu(z_1, \dots,z_n),  \\
_\lambda^{SO(2n+1)}(z_1, \dots , z_n)&\&= 
\sum_{ \ell(\mu)\le n \atop \mu\le |\lambda|} \phi^{SO(2n+1)}_{\lambda \mu}
S_\mu(z_1, \dots,z_n). 
\eea

  This should be compared with Littlewood's character expansion formulae \cite{L,  KT}
\bea
\chi_\lambda^{Sp(2n)} (x_1, \dots, x_n) &\&= 
\sum_{\ell(\mu)\le n \atop \mu\le |\lambda|} \sum_{\alpha}(-1)^{|\alpha|} C^\lambda_{D'(\alpha), \mu} 
S_\mu(x_1, \dots, x_n, x_1^{-1}, \dots , x_n^{-1}) ,
\label{char_SP2n} \\
\chi_\lambda^{SO(2n)} (x_1, \dots, x_n) &\&= 
\sum_{ \ell(\mu)\le n \atop \mu\le |\lambda|} \sum_{\alpha}(-1)^{|\alpha|} C^\lambda_{D(\alpha), \mu} 
S_\mu(x_1, \dots, x_n, x_1^{-1}, \dots , x_n^{-1})  ,
\label{char_SO2n} \\
\chi_\lambda^{SO(2n+1)} (x_1, \dots, x_n) &\&= 
\sum_{ \ell(\mu)\le n \atop \mu\le |\lambda|} \sum_{\alpha}(-1)^{|\alpha|} C^\lambda_{D(\alpha), \mu} 
S_\mu(x_1, \dots, x_n, x_1^{-1}, \dots , x_n^{-1}, 1)
\label{char_SO2n+1} 
\eea
where the sums in  $\alpha$ are  taken over strict partitions
\be 
\alpha = (\alpha_1, \dots \alpha_r), \quad \alpha_i \in \Nb^+,  \quad \alpha_1 > \dots > \alpha_r >0,
\quad 2|\alpha|\le |\lambda|, 
\ee
$D(\alpha)$ and $D'(\alpha)$  are their ``doubles'', defined in Frobenius notation by
\bea 
D(\alpha) &\&:= ( \alpha_1, \dots , \alpha_r |\alpha_1-1, \dots, \alpha_r-1) \cr
D'(\alpha) &\&= ( \alpha_1-1, \dots, \alpha_r-1 | \alpha_1, \dots , \alpha_r),
\eea
and $C^\lambda_{\mu \nu}$ are the Littlewood-Richardson coefficients
 
\subsection{Moment matrices and matrix models.}

Suppose the infinite triangular matrix $\phi_{ij}$ defining our system of polynomials
is chosen to be the one that provides a symmetric lower/upper 
symmetrical factorization of the Hankel matrix of moments 
\be
M_{ij} := \int_\Gamma d\mu(z) z^{i+j}
\ee
of some measure $d\mu$, supported on a curve $\Gamma$ in the complex plane:
\be
\sum_{k =\max{(i,j)}}^\infty \phi_{ki} \phi_{kj} = M_{ij}, \quad i,j \in \Nb.
\ee

It is well known \cite{HO2}  that the chain of  KP $\tau$-functions $\tau_{\phi, \phi} (n, {\bf t}, {\bf 0})$
in (\ref{tau_phi_0}) is then equal to  the sequence of multiple integrals
\bea
\tau_{\phi, \phi}(n, {\bf t}, {\bf 0})  &\& = {1\over n!}\left(\prod_{a=1}^n \int_\Gamma d\mu(z_a) \right)
 \Delta^2({\bf z}) \exp\left(\sum_{j=1}^\infty \sum_{a=1}^n t_j z_a^j\right) \\
 &\& =  \sum_{\ \lambda,\, \ell(\lambda)\le n} B_{\lambda, n}({d\mu}) S_\lambda ({\bf t}) \\
 &\& =  \sum_{\ \lambda,\, \ell(\lambda)\le n}   S^\phi_\lambda ({\bf 0}) S^\phi_\lambda ({\bf t}), 
 \label{1matrix}
\eea
where
\be
\Delta({\bf z}) = \prod_{i<j}^n (z_i -z_j)
\ee
is the Vandermonde determinant and
\be
B_{\lambda, n}(d\mu):= \det(M_{\lambda_i -i + j})_{1\le i,j \le n}
= \sum_{\nu, \, \ell(\nu)\le n}\phi^{(n)}_{\nu \lambda}\phi^{(n)}_{\nu (0)}. 
\ee
 Within a normalization, (\ref{1matrix}) is just  the reduced form of the
 partition function of a generalized conjugation invariant  normal matrix model with
 spectrum supported on $\Gamma$,  expressed as a multiple integral over the eigenvalues \cite{BEH1}.
 
 The matrix of moments can also be expressed in an upper/lower triangular 
 symmetrically factorized form
 \be
M = \tilde{\phi}^{-1} (\tilde{\phi}^t)^{-1},
 \ee
 where the polynomials 
 \be
\tilde{\phi}_i(x) :=\sum_{j=0}^i\tilde{\phi}_{ij}x^j, \quad i\in \Nb
\ee 
  formed from the components $\tilde{\phi}_{ij}$ of the matrix $\tilde{\phi}$
  are orthogonal with respect to the measure $d\mu$
  \be
  \int_\Gamma \tilde{\phi}_i(x) \tilde{\phi}_j(x) d\mu(x)  = \delta_{ij}.
  \ee
  The equality
  \be
 \phi^t\phi = \tilde{\phi}^{-1} (\tilde{\phi}^t)^{-1}
  \ee
  may be viewed as defining a Darboux transformation of the
  measure to the new measure  $d\tilde{\mu}$ whose
  matrix of moments $\tilde{M}$ is
    \be
  \tilde{M}_{ij} = \int_\Gamma d\tilde{\mu} (x) x^{i+j} = \sum_{k=\max(i,j)}^\infty(\tilde{\phi})^{-1}_{ki}(\tilde{\phi})^{-1}_{kj}.
  \ee

\subsection{Bimoments of two variable measures and 2-matrix models.}

We may similarly choose the product of matrices
$\theta^t \phi$ entering in the 2D-Toda chain of $\tau$-functions (\ref{tau_theta_phi})
to be the upper/lower factorization of the matrix of bimoments
\be
M_{ij}:=
\sum_{p,q}c_{pq}\int_{\Gamma_p}\int_{\tilde{\Gamma_q}} d\mu(z,w) z^i w^j = (\theta^t \phi)_{ij}
\label{bimoments}
\ee
of a two-variable measure defined along a  linear combination of products
of curves $\Gamma_p \times \tilde{\Gamma}_q$
in the $z$ and $w$ planes. 

The resulting 2D-Toda $\tau$-function appearing 
in eq.~(\ref{tau_theta_phi}) is then the
partition function of a  coupled two-matrix model  reduced
by a generalized Itzykson-Zuber-Harish-Chandra identity
to a $2n$-fold integral over the eigenvalues {\cite{HO1},
\bea
\tau_{\theta, \phi}(n, {\bf t}, {\bf s}) &\&=
{1\over n!}\prod_{a=1}^n\left( \sum_{p, q}c_{pq}\int_{\Gamma_a}\int_{\tilde{\Gamma_b}} d\mu(z_a,w_a) 
 e^{ \sum_{i=1}^\infty t_i z_a^I} e^{ \sum_{j=1}^\infty s_j w_a^j} \right)\Delta ({\bf z}) \Delta({\bf w})\\
 &\& = \sum_{\lambda\atop\ell(\lambda)\le n} \sum_{\nu \atop \ell(\nu)\le n} B_{\lambda,  \nu, n}(d\nu) S_\lambda({\bf t} )S_\nu ({\bf s})
 \\
 &\& =  \sum_{\lambda \atop\ell(\lambda)\le n} S^\psi_\lambda({\bf t}) S^\phi_\lambda({\bf s}),
\eea
where
\be
 B_{\lambda,  \nu, n}(d\mu) := \det(M_{\lambda_i -i +n, \, \nu_j -j +n})_{1 \le i.j \le n}
 == \sum_{\rho, \, \ell(\rho)\le n}\psi^{(n)}_{\rho \lambda}\phi^{(n)}_{\rho \nu}. 
 \ee

Assuming that all consecutive diagonal $n\times n$ minors of the matrix of bimoments
are nonsingular, we have a system of associated biorthogonal polynomials \cite{BEH2}
\be
\tilde{\phi}_i(x) = \sum_{j=0}^i \tilde{\phi}_{ij}x^j,  \quad \tilde{\theta}_j(y) =\sum_{k=0}^j \tilde{\theta}_{jk} y^k ,
\ee
which may be normalized to have equal leading coefficients 
\be
\tilde{\phi}_{jj} = \tilde{\theta}_{jj}, 
\ee
satisfying
\be
\sum_{p,q}c_{pq}\int_{\Gamma_p}\int_{\tilde{\Gamma_q}} d\mu\tilde{\phi}_i(x) \tilde{\theta}_j(y) = \delta_{ij}.
\ee
It follows that the matrix of moments may be expressed in  factorized lower/upper triangular form as
\be
M = \tilde{\phi}^{-1} (\tilde{\theta}^t)^{-1}
\ee
where $\tilde{\phi}$ and $\tilde{\theta}$ are the lower triangular matrices with 
components $\tilde{\phi}_{ij}$ and $\tilde{\theta}_{ij}$, respectively.
We therefore have the equality
\be
 \tilde{\phi}^{-1}( \tilde{\theta}^t)^{-1} = \theta^t \phi, 
 \ee
 which means that the matrix pair $ (\theta^t, \phi)$ are the Darboux
 transformation of  the pair $(\tilde{\phi}^{-1}, \tilde{\theta}^t{}^{-1} )$,  
 leading to a new measure $d\tilde{\mu}(z,w)$, with bimoment matrix $\tilde{M}$
 \be
 \tilde{M}_{ij} = \sum_{pq}c_{ab}\int_{\Gamma_p}\int_{\tilde{\Gamma_q}} d\tilde{\mu}(z,w)z^iw^j
 = \sum_{k=\max(i,j)}^\infty (\tilde{\theta}{}^{-1} )_{ki} (\tilde{\phi}^{-1} )_{kj}, 
 \ee
which  is thus  the Darboux transformation of the measure $d\mu(z,w)$.

\subsection{Generating function for  nonintersecting random walks}

The unnormalized transition rate for a continuous time,  $n$-particle, right-moving, 
nearest neighbour  exclusion process on the positive integer lattice can be expressed 
as the following fermionic matrix element (cf. \cite{HO4})
\be
\phi^{(n)} _{\lambda \mu}(t) = W_{\mu \ra \lambda}(t, n) = \langle \lambda; n | e^{t \sum_{i=1}^\infty r_i  \psi_i \psi^\dag_{i-1}} | \mu; n\rangle ,
\label{transition_prob}
\ee
which satisfies
\be
{d \phi^{(n)} _{\lambda \mu}(t) \over dt} = \sum_{\nu} M_{\lambda \nu} M_{\nu \mu} ,
\ee
where $\{r_i\}_{i\in \Nb}$ are positive real constants  giving the infinitesimal transition rates
and
\be
 M_{\nu \mu}= \langle \nu; n | \sum_{i=1}^\infty r_i  \psi_i \psi^\dag_{i-1} | \mu; n\rangle.
 \label{M_nu_mu}
\ee

The choice  (\ref{M_nu_mu}) corresponds to a polynomial system  $\{\phi_i\}_{i\in \Nb}$ with coefficient matrix
\be 
\phi = e^\alpha,
\ee
where 
\be
\alpha = \begin{pmatrix} \ddots & \ddots &\vdots  & \cdots & \vdots   \cr
                              \ddots &   r_{j+1}&0  & \cdots & 0   \cr
                                \cdots &   0 & r_j &  \ddots & \vdots  \cr           
                               \cdots &   0 & 0  & \ddots & 0  \cr
                               \cdots &    \vdots & \vdots  & \ddots &r_1  \cr
                               \cdots & 0 & 0 & \cdots   & 0
                                 \end{pmatrix}.
\ee
The associated $\tau$-function (\ref{tau_phi_Schur_exp}) for this case 
\be
\tau_\phi(n, {\bf s}, {\bf t}) = \sum_{\lambda, \mu}  W_{\mu \ra \lambda}(t, n) S_\lambda({\bf s}) S_\mu ({\bf t}), 
\ee
may therefore be viewed as a generating function for the transition probabilities .

Note however that (\ref{M_nu_mu}) is not a normalized transition probability,  since  the  Markov condition
\be
\sum_\nu M_{\nu \mu}= 0
\ee
is not satisfied. To correct this, it would be necessary to add a further term 
 that is quartic in the $\psi$'s and $\psi^\dag$'s.
But  the generating function would then no longer be a $\tau$-function; i.e., this would not 
longer  be a ``free fermion'' process. To obtain a normalized transition
probability in (\ref{transition_prob}), it is therefore necessary to renormalize continuously
at all time values, by dividing by the sum of transition rates to all final states. Alternatively,
the exponential factor in (\ref{transition_prob}) can be developed as a power series in $t$,
with the powers viewed as discrete times. The expansion may then be viewed as a generating function
for a discrete time exclusion process, as in \cite{HO4}, and the normalization
by division of the relative weights by the sum over all final states can be 
applied at each discrete time value.

\medskip

\noindent ==================================================

 \medskip
\noindent {\it Acknowledgements.} The authors would like to thank A. Yu. Orlov for helpful comments 
and assistance with the proof of  Proposition  \ref{g_phi_lambda_mu_prop},    A. Borodin and G. Olshanski
for pointing out ref. \cite{KT} and A.  Veselov for helping to  clarify example 4.1.

\bigskip
\bigskip

\end{document}